\documentclass[prl,twocolumn,showpacs]{revtex4}

\usepackage{graphicx}
\usepackage{amsmath}
\usepackage{epstopdf}
\usepackage{amsmath}
\usepackage{amsfonts}

\usepackage{epsfig}

\newcommand{\Ket}[1]{\left|#1  \right>}

\newcommand{\Braket}[1]{\left<#1  \right>}

\begin{document}

\title{Crossover physics in the non-equilibrium dynamics of quenched quantum impurity systems}

\author{Romain Vasseur$^{1,2}$, Kien Trinh$^{3}$, Stephan Haas${}^3$ and  Hubert Saleur$^{1,3}$}

\affiliation{${}^1$Institut de Physique Th\'eorique, CEA Saclay,
91191 Gif Sur Yvette, France}
\affiliation{${}^2$LPTENS, 24 rue Lhomond, 75231 Paris, France}
\affiliation{${}^3$Department of Physics and Astronomy,
University of Southern California, Los Angeles, CA 90089-0484}

\date{\today}

\begin{abstract}

A general framework is proposed to tackle analytically local quantum quenches in integrable  impurity systems,  combining a mapping onto a boundary problem with the form factor approach to boundary-condition-changing operators introduced in  Phys. Rev. Lett. 80, 4370 (1998). We discuss how to compute exactly two central quantities of interest: the Loschmidt echo and the distribution of the work done during the quantum quench. Our results display an interesting crossover physics characterized by the energy scale $T_b$ of the impurity corresponding to the Kondo temperature. We discuss in detail the non-interacting case as a paradigm and benchmark for more complicated integrable impurity models, and check our results using numerical methods.

\end{abstract}


\maketitle

\paragraph{Introduction.}

Quenches are a clear-cut way to study the non-equilibrium physics of quantum systems and reveal their intrinsic time scales.  Here, energy is injected into an otherwise closed system at time $t=0$ via a sudden change of a control parameter, and it is subsequently left to evolve unitarily. From a theoretical perspective this is convenient because energy exchange between the system and its environment is not considered during its evolution. Furthermore, such quench scenarios can be realized by experiments probing the transient dynamics of physical systems, such as atoms in a tunable trap potential or quantum dots connected to tunable control gates. 

In many cases, the effect of interactions is to give rise to crossovers between two different kinds of physical behavior. In the Kondo problem for instance, the crossover takes place between  a weakly coupled two level system (the spin $1/2$ impurity) at high energy, and a strongly coupled screened impurity at low energy, and is characterized by the Kondo temperature $T_K$~\cite{Hewson}. This crossover has been widely studied in equilibrium~\cite{Hewson}. It has also been considered in non equilibrium situations, both experimentally~\cite{Exp1,Exp2} and numerically~\cite{HKCrooks,HKXray}. 

Many theoretical methods have been developed to attack quantum impurity problems analytically~\cite{AndreiLowenstein,Affleckreview}, and it is reasonable to expect that progress can be made in the study of quench dynamics as well. Note however that the most potent method - the use of conformal invariance~\cite{CardyCalabrese} - does not apply in the case of crossovers, since these precisely describe the departure from scale invariance. This means that in quantum impurity problems, only the very long time, low energy, behavior can be described by methods such as those in~\cite{CardyCalabrese,WorkSilva}. 

Of all the quantities one may want to study in quenches, the most fundamental is probably the overlap of the states before and after the quench~\cite{Delft}. In equilibrium, this is deeply related to the Anderson orthogonality catastrophe~\cite{Anderson}.  The type of situation we have in mind here is rather the work distribution in a situation where an impurity is suddenly coupled to an electron bath, the coupling being characterized by a crossover temperature $T_b$ in equilibrium. We expect that the work distribution should be a function of $W/T_b$, barring some scaling violations such as those observed in the screening cloud problem~\cite{AffleckBarzykin}. 

We discuss here the main ideas of our approach, and illustrate them in the case of the resonant level model (RLM). The Kondo model can be discussed similarly, at the price of some extra technical difficulties that will be presented elsewhere. We immediately warn the reader that, although the RLM is known as a `non-interacting model', the quench is an extremely non-trivial operation in the free electron basis, as it affects an infinity of multiple particle states. The technical difficulties involved in its study are entirely analogous to those occurring in the Kondo case. 

General quantum mechanics considerations show that the Loschmidt echo and the work distribution function can be expressed in terms of the scalar products of eigenstates of the Hamiltonian before and after the quench. The main point of our approach is that these scalar products can all be determined analytically, in the limit where all energy scales are  much smaller than the cutoff (bandwidth),  by using an axiomatic formulation very much like what is done in the form-factor approach to correlations in integrable massive quantum field theories. The corresponding theoretical tools were put forward in earlier papers~\cite{BCCFF1,BCCFF2}. There are however several major difficulties. First, an infinity of such matrix elements has to be taken into account. Second, the associated sums over eigenstates are strongly divergent at low energies, via an infrared catastrophe inherent to the fact that we are dealing with a massless bulk theory. As we shall see, these difficulties can be controlled using  renormalization tricks, and remarkably accurate results from the low to the high energy regions can be obtained. 

\paragraph{Resonant Level Model.}

\begin{figure}[t!]
\includegraphics[width=1.0\linewidth]{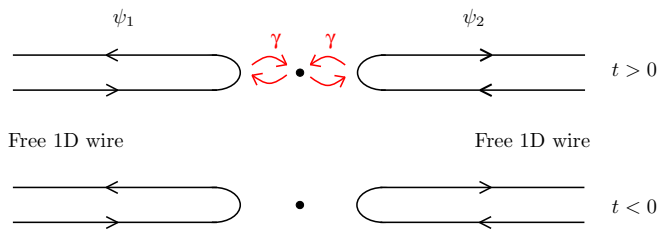}
\caption{Quantum quench in the Resonant Level Model.}
\label{fig1}
\end{figure}

The spinless RLM involves  two independent one-dimensional wires connected by tunneling  through a quantum dot (the `impurity'). The RLM is equivalent to the anisotropic Kondo system at the Toulouse point, and to the problem of an impurity in a Luttinger liquid with parameter $g=\frac{1}{2}$, and therefore the approach outlined below will apply to these systems as well.
After unfolding the wires, the Hamiltonian reads $ H(\gamma) =  \int {\rm d} x \mathcal {H}(\gamma)$, with the Hamiltonian density
\begin{equation}
\mathcal {H}(\gamma) = - i v_F \sum_{a=1,2} \psi_a^\dagger \partial_x \psi_a + \delta(x) \frac{\gamma}{\sqrt{2}} \sum_a \psi_a^\dagger(0) d + h.c.
\label{eqRLM_H}
\end{equation}
Here, the label $a$ denotes the two wires, $\gamma$ is a tunneling amplitude (which we took, without loss of generality, to be the same for both wires). The equilibrium physics of the RLM is rather simple. It is convenient to define $\psi_{\pm} = \frac{1}{\sqrt{2}}(\psi_1 \pm \psi_2)$, so that  $\psi_{-}$ decouples from the impurity. The scattering matrix of the remaining fermion $\psi_{+}$ on the impurity then reads $S_{+}(\omega) = \frac{i \omega -T_b}{i \omega +T_b}$. The tunneling term is a relevant interaction, thus creating an energy scale $T_b = \frac{\gamma^2}{2}$, and the system flows under renormalization from the $\gamma=0$ fixed point (independent wires) to a strong coupling fixed point $\gamma = \infty$ where the impurity is completely hybridized with the wires. At low energy, the only remaining effect of the impurity is a phase shift $\psi_{+}(0^+) = \mathrm{e}^{2 i \delta} \psi_{+}(0^-)$ with $\mathrm{e}^{2 i \delta} = S_{+}(\omega=0)$ so $\delta = \frac{\pi}{2}$. 

\paragraph{Quantum quench.}
The main goal of this paper is to understand analytically the quantum dynamics of this system after suddenly turning on (or off) the tunneling $\gamma$. Let $H_0 = H(\gamma=0)$ be the Hamiltonian of the system for $t<0$, and $H_1=H(\gamma)$ the Hamiltonian for $t \geq 0$ (see Fig.~\ref{fig1}). The framework presented here is quite general and can be applied at finite temperature, but for simplicity, we will only consider the case $T=0$ and imagine that the system is initially prepared in the groundstate $\Ket{\Psi(0)}=\Ket{\psi^{(0)}_{0}}$ of $H_0$ for $t<0$. The wave function of the system at time $t$ is then $\Ket{\Psi(t)} = \mathrm{e}^{-i H_1 t} \Ket{\Psi(0)}$.

\paragraph{Work distribution and crossovers.} 
We are interested  in the work performed on the system during the quantum quench. In order to determine the work done $W$, two energy measurements are necessary, so that the work is not an observable but is rather characterized by a distribution function, $ P(W) = \sum_n \left| \Braket{\psi^{(n)}_{1} \left|\psi^{(0)}_{0}} \right|^2 \right. \delta(W-(E^{(n)}_1-E^{(0)}_0))$, where $\psi^{(n)}_{0,1}$ and $E^{(n)}_{0,1}$ are the eigenstates and energies of the Hamiltonian $H_0$ before (resp.  $H_1$ after) the quench. The work distribution $P(W)$ has gathered a lot of attention recently~\cite{WorkSilva, HKCrooks,HKXray,Goold1,Gambassi}, and is especially of interest since it can be measured experimentally by spectroscopy~\cite{Exp1,Exp2,HKCrooks} (see also~\cite{Goold2}). The generating function of the moments of the work $G(t) = \int {\rm d} W \mathrm{e}^{-i W t} P(W)$ is given~\cite{WorkSilva,Campos} by the Loschmidt echo $G(t) = \Braket{\psi^{(0)}_{0} |\mathrm{e}^{i H_0 t}  \mathrm{e}^{-i H_1 t}|\psi^{(0)}_{0}}$, which is, up to a phase, nothing but the overlap $\Braket{\Psi(t)|\Psi(0)}$. For  energies much smaller than the cutoff (bandwidth), the Loschmidt echo and the work distribution take the universal forms $G(t)=g(t T_b)$, and $P(W)=\frac{1}{T_b}p(\frac{W-\delta E}{T_b})$, where $\delta E=E^{(0)}_{1}-E^{(0)}_{0}$ is the minimal work that should be performed on the system during the quench. They show some interesting crossover dynamics at $t \sim T_b^{-1}$ ($W-\delta E \sim T_b$) (here, $T_b$ is the equivalent of $T_K$). This contrasts with the case of gapless quantum spin chains for which the Loschmidt echo at low energy is a pure power-law with an exponent related to the central charge~\cite{EchoCFT1,EchoCFT2}. The time evolution after the quench follows the Renormalization Group (RG) flow all the way from the UV ($t \ll T_b^{-1} $, weak coupling) to the IR low energy limit ($t \gg T_b^{-1} $, strong coupling). The computation of a quantity such as $G(t)$ thus requires all the non-perturbative physics in order to observe the crossover at $t \sim T_b^{-1}$. Note that even in a free-fermion problem such as the RLM, the computation of $G(t)$ is highly non-trivial because it is non-local in terms of the fermions.

\begin{figure}[t!]
\includegraphics[width=0.9\linewidth]{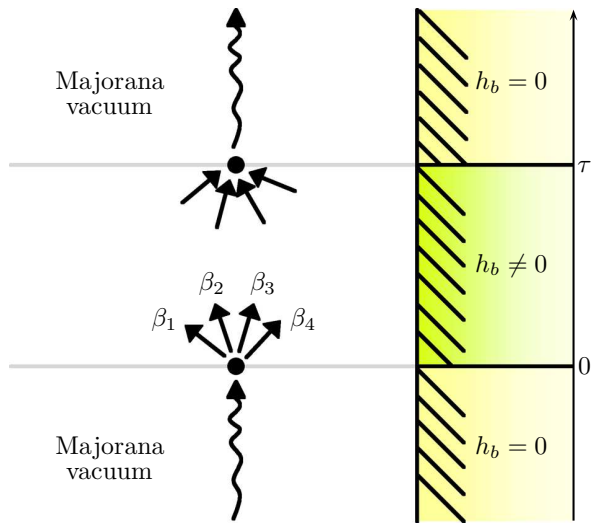}
\caption{The Loschmidt echo in the resonant level model can be thought of as a partition in the Ising model with alternating vanishing and finite boundary magnetic fields $h_b$. The quantum quench then effectively creates an arbitrary number of fermionic  excitations in the system. $\tau$ is the imaginary time.}
\label{fig2}
\end{figure}

\paragraph{Mapping onto a boundary problem.} In order to compute the Loschmidt echo $G(t)$, we first map the quantum impurity system onto an equivalent boundary problem. This can be done in general, and the Loschmidt echo $G(t)$ can then be interpreted as the partition function of a 2D statistical problem, critical in the bulk, with non-conformally invariant boundary conditions. In our example of the RLM, after folding the $\psi_{+}$-fermions system and decomposing $\psi_{+}$ into Majorana (real) components, one finds that the RLM is equivalent to two independent copies of the critical Ising model, with a boundary magnetic field $h_b = \frac{\gamma}{\sqrt{2}}$. This is consistent with the fact that the scattering matrix $S_{+}(\omega)$ has the form of the reflection matrix in the Ising field theory~\cite{GZ}. The Loschmidt echo after a Wick rotation then reads $G(t=-i \tau) = (Z_{h_b}(\tau))^2$, where $Z_{h_b}(\tau)$ is the partition function of the critical Ising model in the half-plane, with finite boundary magnetic field $h_b$ for imaginary times between $0$ and $\tau$, and free boundary conditions (BC) elsewhere (see Fig.~\ref{fig2}). Therefore, the Loschmidt echo following  a quench in an integrable impurity problem reduces to the computation of a partition function in an integrable classical system with alternating zero and finite boundary magnetic fields. Note also that $Z_{h_b}(\tau)$ is normalized so that $Z_{h_b}(\tau=0)=Z_{h_b=0}(\tau)=1$. 

\paragraph{Low energy limit.} The computation of $Z_{h_b}(\tau)$ is non-trivial because the non-zero  boundary magnetic field creates an energy scale $T_b = h_b^2$, and thus the problem cannot be solved using Boundary Conformal Field Theory (BCFT) techniques. However, in the low-energy limit $\tau h_b^2 \gg 1$, the BC becomes conformally invariant and $Z_{h_b}(\tau)$ coincides with the two-point function of the Boundary Conditions Changing (BCC) operator from free to fixed BC in the Ising model. The scaling dimension of the latter is known to be $h = \frac{1}{16}$, so that $Z_{h_b}(\tau) \sim (h_b \tau)^{-1/8}$ in that limit. Going back to the Loschmidt echo, this means that $G(t) \sim t^{-1/4}$ when $t T_b \gg 1$, where the exponent can also be interpreted in terms of the phase shift of the $\psi_{+}$ fermions $2 h = \frac{1}{2}\left(\frac{\delta}{\pi} \right)^2 =\frac{1}{8}$. The vanishing of $G(t)$ can therefore be traced back to the well-known Anderson orthogonality catastrophe. In terms of the work distribution, we thus obtain an edge singularity at low energy~\cite{WorkSilva, HKXray} 
\begin{equation}
P(W)  \underset{W-\delta E \ll T_b}{\propto} \frac{1}{T_b} \theta(W-\delta E) \left(\frac{W-\delta E}{T_b}\right)^{\alpha-1},
\label{eqEdgesingularity}
\end{equation}
with $\alpha = \left(\delta / \pi \right)^2 = 1/4$ in the RLM. This conclusion holds also for quenches in interacting impurity models, and one finds for example $\alpha = g/2$ for the anisotropic Kondo case ($g=1$ being the isotropic model, and $g=1/2$ the Toulouse point); and $\alpha = 1/8$ for an impurity in a Luttinger liquid, regardless of the Luttinger parameter.

\paragraph{Form Factor approach.} Although BCFT techniques can be used to analyze the low-energy limit of the quantum quench, the full non-perturbative computation of $G(t)$ is much more involved. When $h_b$ is finite, one can still think of $Z_{h_b}(\tau)$ as the two-point function of some generalization of BCC operators to non-conformal field theory. For integrable systems such as the ones we are considering, it should come as no surprise that the matrix elements of these operators can be handled~\cite{BCCFF1,BCCFF2} using axiomatic techniques very similar to those used in the  Form Factor (FF) approach to bulk correlations~\cite{Smirnov}. In particular, we find that the quench of the boundary magnetic fields creates an arbitrary number of fermions in the Ising field theory (see  Fig.~\ref{fig2}), so that one obtains the formal expansion 
\begin{equation}
Z_{h_b}(\tau) = \sum_{n=0}^{\infty} \int \prod_{i=1}^{n} \frac{{\rm d} \beta_i}{2\pi} \frac{1}{n!}\mathrm{e}^{-\tau(\sum_i \mathrm{e}^{\beta_i})} |~_{h_b} \Braket{\beta_1,\dots,\beta_n | \Omega}_0 |^2, \label{ffexp}
\end{equation}
where $\beta_i$ is the rapidity of the $i^{\rm th}$ fermionic excitation of energy $\mathrm{e}^{\beta_i}$. $~_{h_b} \Braket{\beta_1,\dots,\beta_n | \Omega }_0$ is the overlap between the groundstate of the Ising model $\Ket{\Omega}_0$ with free boundary conditions, and the eigenstates $\Ket{\beta_1,\dots,\beta_n}_{h_b}$ of the model in the presence of a finite boundary magnetic field $h_b$. This inner product can be thought of as a FF of BCC operator in the boundary Ising model, which  is  {\it known exactly} from the  axiomatic approach~\cite{BCCFF1,BCCFF2}. (For other works using integrability  ideas in the context of quenches, see ~\cite{qu1,qu2,qu3}).

\begin{figure*}
\centering
\begin{tabular}{cc}
\epsfig{file=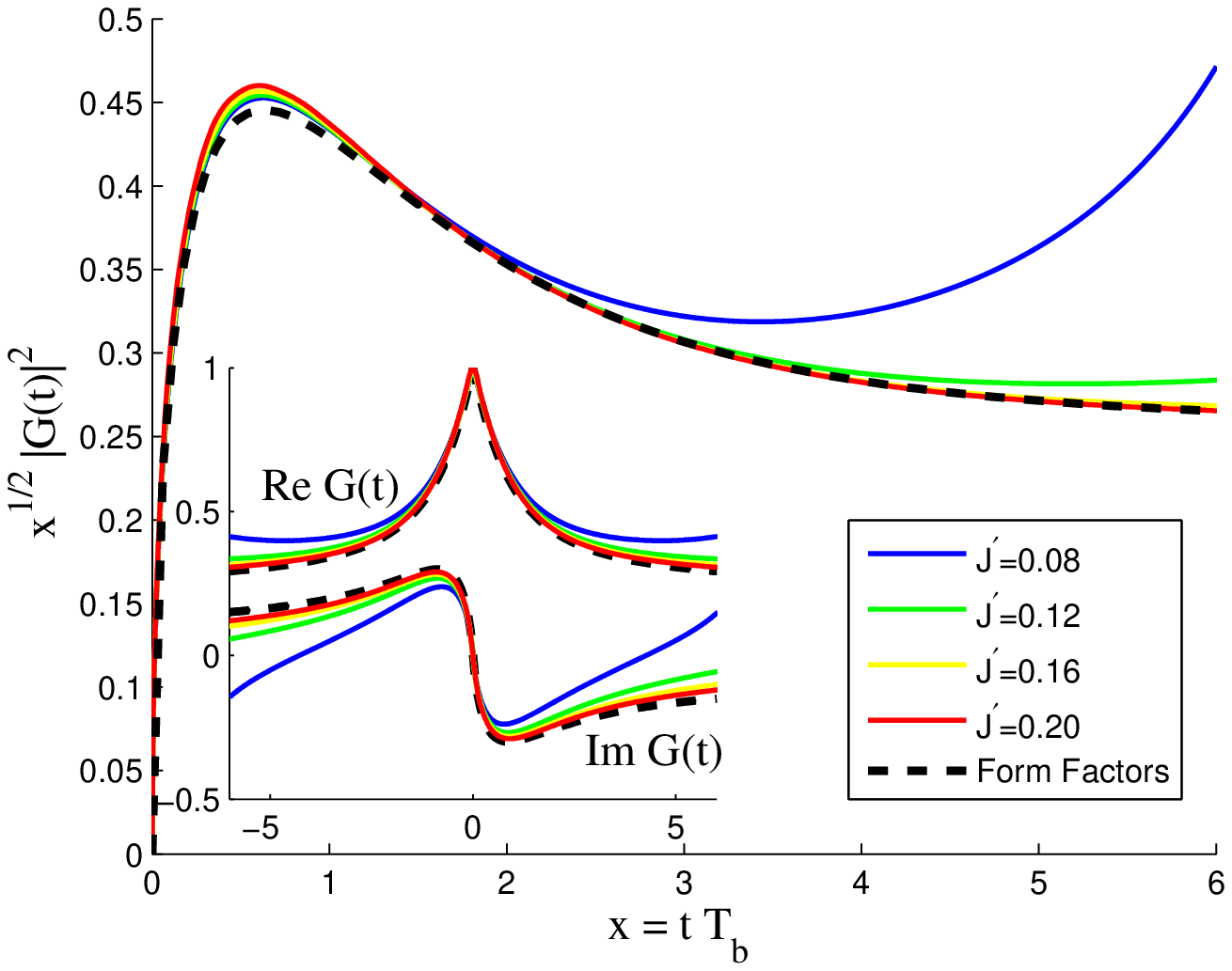,width=0.5\linewidth,clip=} &
\epsfig{file=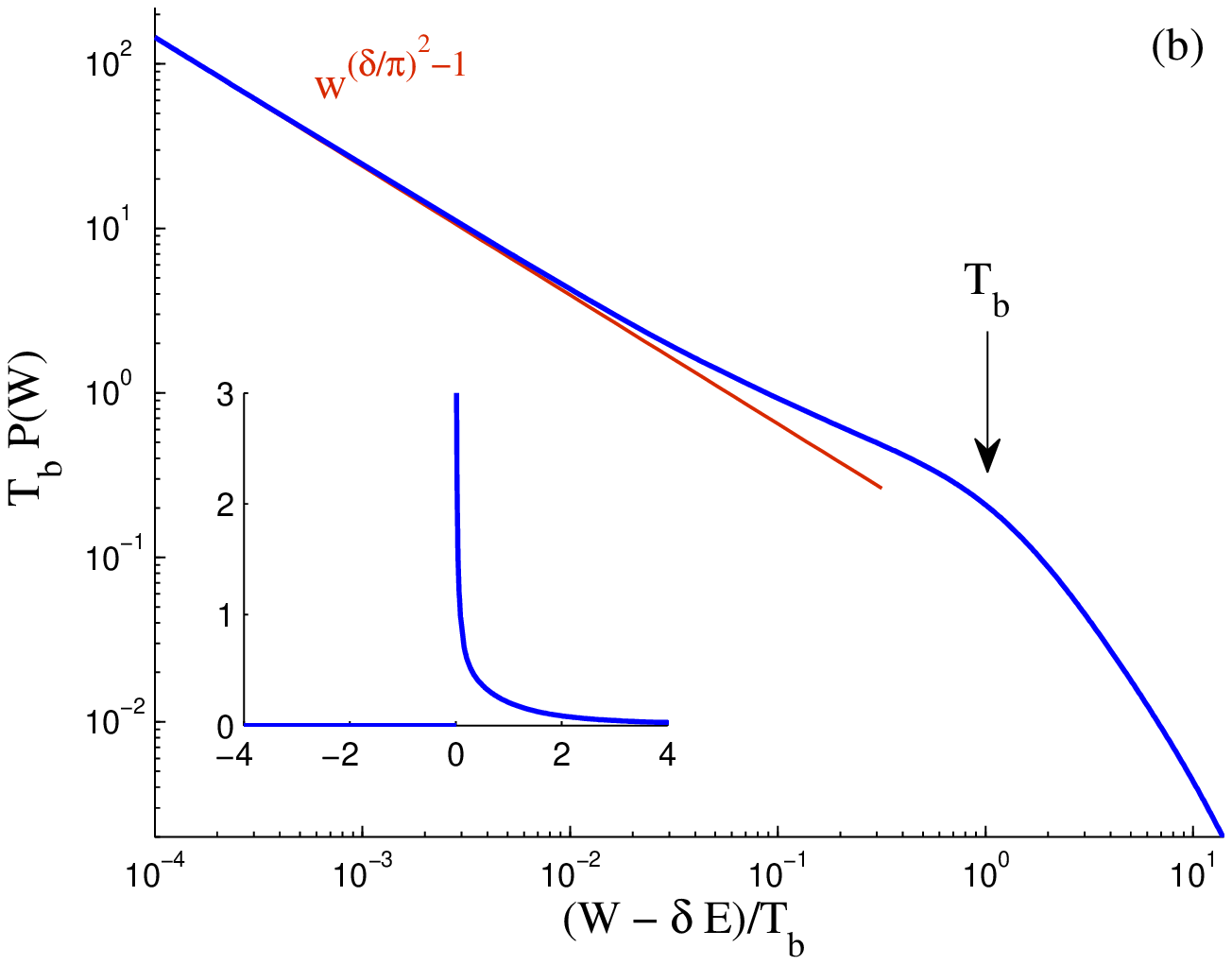,width=0.5\linewidth,clip=} 
\end{tabular}
\caption{(a) Loschmidt echo as a function of $x=t T_b$, comparison between numerical results and Form Factors. Note that we rescaled the Loschmidt echo $|G(t)|^2$ in the main figure by a factor $x^{1/2}$ in order to cancel the asymptotic power-law behavior for $x= t T_b \gg 1$. (b) Universal work distribution from FF. At low energy, the work distribution has an edge singularity with an exponent $\alpha = \left(\frac{\delta}{\pi} \right)^2 =\frac{1}{4}$ given by the dimension of a BCC operator while for $W-\delta E \sim T_b$, the system shows an interesting crossover physics.}
\label{FigResults}
\end{figure*}

Now, expression (\ref{ffexp}) involves an infinite sum of terms. Since the problem is massless in the bulk, $n$ particle terms do not necessarily give rise to smaller contributions, as their energy can be arbitrarily low. The convergence of the expansion (\ref{ffexp}), or even its numerical usefulness, are thus not obvious a priori. Moreover, the integrals involved in this expansion are in fact IR divergent!  It turns out nevertheless that the sum  can be regularized by reorganizing the terms in the expansion. Details will be given elsewhere. We only quote  the leading-order  final result,
\begin{multline*}
\ln Z_{h_b}(\tau) = \int_{0}^{\infty} \frac{{\rm d} u}{2 \pi u} g_x(u) \Psi(u)  + \frac{1}{2!} \int_{0}^\infty \frac{{\rm d} u_1}{2 \pi u_1}  \\ \int_{0}^\infty \frac{{\rm d} u_2}{2 \pi u_2} g_x(u_1+u_2) \left( \left(\frac{u_1-u_2}{u_1+u_2}\right)^2 - 1 \right) \Psi(u_1) \Psi(u_2) +\dots ,
\end{multline*}	
where $x= \tau T_b$ (recall $\tau$ is the imaginary time) is an effective   coupling parameter, dimensionless as expected, $g_x(u)= \mathrm{e}^{-x u}-1$, and the Kernel $\Psi(u)$ reads
\begin{equation}
\Psi(u)= \frac{\sqrt{u}}{1+u^2} \exp \left[ \int_{- \infty}^{\infty} \frac{{\rm d} t}{2t} \left( \frac{2}{t} - \frac{\cos \frac{\ln u}{2 \pi} t}{\cosh \frac{t}{4} \sinh \frac{t}{2}} \right) \right],
\end{equation}
with $\Psi(0)=1$ and $\Psi(u) \sim u^{-1}$ as $u \rightarrow \infty$.
As a non-trivial check, this expansion can be resummed exactly in the IR limit $x \gg 1$, which yields $Z_{h_b}(\tau) \sim x^{-1/8}$ as expected from CFT. Moreover, for any value of $x$, $Z_{h_b}(\tau)$ can be estimated by keeping only a few terms in the FF expansion. Typically, the two and three-particle contributions are sufficient to obtain a very good approximation all over the crossover region. Finally, we note that, although we have used imaginary time to allow for a  statistical mechanics interpretation of the results, everything works as well in real time, with $\tau=it$.

\paragraph{Lattice model.}

In order to validate the FF approach, we compare our analytical results to numerics in the RLM case~\eqref{eqRLM_H}. We consider the itinerant fermion model (equivalent to a XX spin-1/2 chain) with two weak links,
\begin{equation}
H = - J \sum_{i} (c^\dagger_{i+1} c_i + h.c.) - (J^\prime - J) (c^\dagger_{0} c_{-1} +c^\dagger_{1} c_{0} + h.c.),
\label{eqLattice}
\end{equation}
with $J=1$ so that the Fermi velocity is $v_F=2$. At sufficiently low energies $J^\prime \ll J=1$, the system is described by the effective Hamiltonian~\eqref{eqRLM_H},  with $\gamma\propto J'$. Given the free fermionic nature of the problem, it is even possible to identify exactly the energy scale $T_b \simeq 2 J^{\prime 2}/J$~\cite{TbRef}. The Loschmidt echo $G(t)$ following a quench from $J'=0$ to $J'\neq 0$ can be expressed as a determinant that we evaluate numerically -- see~\cite{EchoCFT2} for similar calculations. We compute $G(t)$ on $L=1025$ sites for different values of $J^\prime$, and find that the results indeed collapse onto a universal curve after rescaling  the time scale by a factor $T_b$. 



The universal curve of the Loschmidt echo computed from numerics and from an extrapolated (see supplementary material) two-particle FF expansion are shown in Fig.~\ref{FigResults}-(a). Note that because of finite size effects, one expects the curves for small values of $J^\prime$ to describe well the universal curve for small $x$ only. We find that the FF expansion is in very good agreement with our numerical results, even in the interesting non-perturbative region $t \sim T_b^{-1}$ where $G(t)$ has a non-trivial behavior -- note that there is no free parameter in the results, which must match without possible rescaling of the time axis. The resulting universal work distribution is shown in Fig.~\ref{FigResults}-(b). As expected, we observe an edge singularity~\eqref{eqEdgesingularity} at low energy. Note that this power-law singularity would be smeared at finite temperature. For $W \sim  \delta E + T_b$, the work distribution has a bump which we interpret as a signature of a Kondo resonance. We emphasize that although some aspects of our work are well-known in equilibrium (Anderson orthogonality catastrophe, crossover temperature $T_b$), it is truly remarkable to observe this `Kondo physics' in the real-time dynamics of the system.

\paragraph{Discussion.} 

The new non-equilibrium dynamics approach presented here is based on a generalization of the form-factor approach. The calculated time evolution of the Loschmidt echo was found to be in beautiful agreement with  independent numerical studies of an equivalent lattice model. Applying the same formalism to the the Kondo case is a bit more involved, since the form factors are more complicated in this case. However,  the problem is not fundamentally different from the case of the RLM, the physics of the quenches being essentially interacting in that case as well.

Crossovers in quench dynamics are just beginning to get studied experimentally, especially in the context of quantum impurity problems. Recently, in Ref.~\cite{Exp2} a Kondo crossover scale in the absorption energy of a quenched quantum dot system was  reported, reminiscent of our results in the time domain (Fig.~\ref{FigResults}-(b)). One could well imagine analogous experiments in the context of cold atom systems, whereby a local quench is applied to the trapping potential, and the subsequent time evolution of the zero-momentum occupation number is monitored for associated crossover effects.

\smallskip

\paragraph{Acknowledgments.}
We wish to acknowledge useful discussions with Natan Andrei, Lorenzo Campos Venuti and Paolo Zanardi.  This work was supported by the the French Agence Nationale pour la Recherche (ANR Projet 2010 Blanc SIMI 4 : DIME) and the US Department of Energy (grant number DE-FG03-01ER45908).

\vfill
\eject

\clearpage
\newpage

\centerline{\bf Supplementary material}

\bigskip

In this supplementary material, we provide more details regarding the FF calculations. The problem is to compute the Ising matrix elements $|\;_{h_b}\Braket{\beta_1,\dots,\beta_n | \Omega }_0 |^2$ in the expansion
\begin{align}\label{eqAp1}
Z_{h_b}(\tau) &= \sum_{n=0}^{\infty} \int \prod_{i=1}^{n} \frac{{\rm d} \beta_i}{2\pi} \frac{1}{n!}\mathrm{e}^{-\tau(\sum_i \mathrm{e}^{\beta_i})} \notag \\
& \times |\;_{h_b}\Braket{\beta_1,\dots,\beta_n | \Omega }_0 |^2.
\end{align}
Those matrix elements were computed exactly for an arbitrary change of BC from fixed boundary magnetic field $h_a$ to $h_b$ in~\cite{BCCFF1,BCCFF2}. In the case $h_a h_b >0$, fermions are created by pairs, whereas the case $h_a h_b <0$ requires an additional particle to account for the presence of a domain wall. The intermediate case we consider here is  $h_a h_b=0$ (free BC corresponds to $h_a=0$), and it  allows for an arbitrary number of fermions. We will start from the case $h_a h_b >0$,  the FF of the corresponding BCC operator read
\begin{widetext}
\begin{equation}
|G_{ba}(\beta_1, \dots, \beta_{2n})|^2 \equiv   \frac{|\;_{h_b} \Braket{\beta_{2n},\dots,\beta_1| \Omega}_{h_a}|^2 }{|\;_{h_b}\Braket{\Omega|\Omega}_{h_a} |^2} =  \frac{1}{4^n} \prod_i \sinh \frac{\beta_b-\beta_a}{2} \left|\Phi(\beta_i) \right|^2 \prod_{i<j} \tanh^2 \frac{\beta_i-\beta_j}{2} , 
\end{equation}
\end{widetext}
where $\Ket{\Omega}_{h}$ is the vacuum of the theory with boundary magnetic field $h$, $T_a=h_a^2=\mathrm{e}^{\beta_a}$, $T_b=h_b^2=\mathrm{e}^{\beta_b}$, and $\beta_i$ is the rapidity of the $i^{\rm th}$ particle. The function $\left|\Phi(\beta_i) \right|^2$ has the following integral representation 
\begin{multline}
\Phi(\beta|\beta_a, \beta_b) =  \frac{1}{\cosh \left( \frac{\beta-\beta_b}{2}-\frac{i \pi}{4}\right)} \\
\times \exp \left[ \int_{-\infty}^{+\infty} \frac{{\rm d} t}{t} \dfrac{\cos(\frac{\beta-\beta_a}{2 \pi} t) - \cos(\frac{\beta-\beta_b}{2 \pi} t)}{4 \cosh \frac{t}{4} \sinh \frac{t}{2}}\right].
\end{multline}
In order to describe free boundary conditions, we should take the limit $\beta_a \to - \infty$. We find 
\begin{equation}
|\Phi(\beta|\beta_a, \beta_b)|^2 \underset{h_a \to 0}{\sim} \mathrm{e}^{-(\beta_b - \beta_a)/2} 4 \Psi \left(\frac{u}{T_b} \right),  
\end{equation}
with $u=\mathrm{e}^\beta$, and $\Psi(u)$ is the kernel used in the main text
\begin{equation}
\Psi(u)= \frac{\sqrt{u}}{1+u^2} \exp \left[ \int_{- \infty}^{\infty} \frac{{\rm d} t}{2t} \left( \frac{2}{t} - \frac{\cos \frac{\ln u}{2 \pi} t}{\cosh \frac{t}{4} \sinh \frac{t}{2}} \right) \right].
\end{equation}
Using this function, the FF for our case $h_a h_b=0$ can be expressed as
\begin{multline}\label{eqApFF}
|G_{b}(\beta_1, \dots, \beta_{n})|^2 \equiv   \frac{|\;_{h_b} \Braket{\beta_{n},\dots,\beta_1| \Omega}_0|^2 }{|\;_{h_b}\Braket{\Omega|\Omega}_0 |^2} \\
=  \prod_i \Psi \left(\frac{u_i}{T_b} \right) \prod_{i<j} \left( \frac{u_i-u_j}{u_i+u_j}\right)^2 , 
\end{multline}
where $u_i=\mathrm{e}^{\beta_i}$. We also recall that $\Ket{\Omega}_0$ is the vacuum of the theory for free BC. In terms of these FF, the expansion~\eqref{eqAp1} becomes
\begin{equation}\label{eqAp2}
Z_{h_b}(\tau) = \frac{\sum_{n=0}^{\infty} \int \prod_{i} \frac{{\rm d} \beta_i}{2\pi} \frac{1}{n!}\mathrm{e}^{-\tau(\sum_i \mathrm{e}^{\beta_i})} |G_{b}(\beta_1, \dots, \beta_{n})|^2}{\sum_{n=0}^{\infty} \int \prod_{i} \frac{{\rm d} \beta_i}{2\pi} \frac{1}{n!} |G_{b}(\beta_1, \dots, \beta_{n})|^2} .
\end{equation}
If we plug~\eqref{eqApFF} into this expression, we find that the integrals diverge at low energy $u_i=\mathrm{e}^{\beta_i} \to 0$. However, the ratio can be regularized via a formal expansion of the inverse denominator and proper regrouping of the terms. This is similar to the trick used in~\cite{LesageSaleurFriedel}. More precisely, we find, introducing a small IR cutoff $\Lambda$, 
\begin{widetext}
\begin{equation}\label{eqAp3}
\ln Z_{h_b}(\tau) = \ln |\;_{h_b}\Braket{\Omega|\Omega}_0 |^2 + \int_{\Lambda}^{\infty} \frac{{\rm d} u}{2 \pi u} \mathrm{e}^{-\tau u}  \Psi \left(\frac{u}{T_b} \right) +  \int_{\Lambda}^\infty \frac{{\rm d} u_1 {\rm d} u_2}{2! (2 \pi)^2 u_1 u_2}  \mathrm{e}^{- \tau (u_1+u_2)}  \left( \left(\frac{u_1-u_2}{u_1+u_2}\right)^2 - 1 \right) \Psi\left(\frac{u_1}{T_b} \right) \Psi \left(\frac{u_2}{T_b} \right) +\dots
\end{equation}
\end{widetext}
When $\Lambda \to 0$, the integrals in this expression are divergent, and this is related to the Anderson catastrophe as at low energy $|\;_{h_b}\Braket{\Omega|\Omega}_0 |^2 \sim |\;_{\infty}\Braket{\Omega|\Omega}_0 |^2 \sim 0$, so the first term $ \ln |\;_{h_b}\Braket{\Omega|\Omega}_0 |^2$ is also infinite. To regularize, we subtract the value at $\tau=0$
\begin{equation}\label{eqAp4}
\ln Z_{h_b}(\tau=0) = 0  = \ln |\;_{h_b}\Braket{\Omega|\Omega}_0 |^2 + \int_{\Lambda}^{\infty} \frac{{\rm d} u}{2 \pi u}  \Psi \left(\frac{u}{T_b} \right) + \dots
\end{equation}
Replacing the value of $  \ln |\;_{h_b}\Braket{\Omega|\Omega}_0 |^2$ into~\eqref{eqAp3}, we can now take the cutoff to zero $\Lambda \to 0$ and we find the well-defined expression
\begin{widetext}
\begin{multline}
\ln Z_{h_b}(\tau) = \int_{0}^{\infty} \frac{{\rm d} u}{2 \pi u} (\mathrm{e}^{-x u}-1)  \Psi(u)  + \frac{1}{2!} \int_{0}^\infty \frac{{\rm d} u_1}{2 \pi u_1}  \int_{0}^\infty \frac{{\rm d} u_2}{2 \pi u_2} (\mathrm{e}^{-x (u_1+u_2)} - 1)  \left( \left(\frac{u_1-u_2}{u_1+u_2}\right)^2 - 1 \right) \Psi(u_1) \Psi(u_2) \\ + \frac{1}{3!} \int_{0}^\infty \frac{{\rm d} u_1}{2 \pi u_1}  \int_{0}^\infty \frac{{\rm d} u_2}{2 \pi u_2}  \int_{0}^\infty \frac{{\rm d} u_3}{2 \pi u_3}   (\mathrm{e}^{-x (u_1+u_2+u_3)} - 1)  \left[ \left(\frac{u_1-u_2}{u_1+u_2}\right)^2 \left(\frac{u_1-u_3}{u_1+u_3}\right)^2 \left(\frac{u_2-u_3}{u_2+u_3}\right)^2 +2 \right. \\
\left. - \left(\frac{u_1-u_2}{u_1+u_2}\right)^2 -\left(\frac{u_1-u_3}{u_1+u_3}\right)^2 -\left(\frac{u_2-u_3}{u_2+u_3}\right)^2\right]  \Psi(u_1) \Psi(u_2) \Psi(u_3) +\dots
\end{multline}
\end{widetext}
where we recall that $x=\tau T_b$. We emphasize that these IR divergences can be traced back to the Anderson orthogonality catastrophe as for $h_b \to \infty$, the inner product $~_\infty\langle\Omega|\Omega\rangle_0$ should be {\sl exactly zero} in the thermodynamic limit, so~\eqref{eqApFF} becomes infinite. 

The regularized integrals in the FF are then computed numerically by keeping only a few terms (2 or 3 typically) in the expansion. The FF expansion is alternating,  and averaging the expressions obtained at first and second orders has been shown to increase the precision by more than an order magnitude in the case where the analytical solution is known~\cite{LesageSaleurFriedel}. We obtain in this way an excellent approximation in the present case as well.


\begin{thebibliography}{99}



\bibitem{Hewson} A. Hewson, The Kondo problem to heavy fermions, Cambridge studies in Magnetism, Cambridge 

\bibitem{Exp1} H.E. T\"ureci {\it et al}, Phys. Rev. Lett. {\bf 106}, 107402 (2011).

\bibitem{Exp2} C. Latta {\it et al}, Nature {\bf 474}, 627--630 (2011).

\bibitem{HKCrooks} M. Heyl and S. Kehrein, Phys. Rev. Lett. {\bf 108}, 190601 (2012).

\bibitem{HKXray} M. Heyl and S. Kehrein, Phys. Rev. B {\bf 85}, 155413 (2012).

\bibitem{AndreiLowenstein} N. Andrei, K. Furuya and F. Lowenstein,  Rev. Mod. Phys. {\bf 55}, 331 (1983).

\bibitem{Affleckreview} I. Affleck,  Conformal field theory approach to quantum impurity problems, {\tt arxiv:cond-mat/9311054}.

\bibitem{CardyCalabrese}  P. Calabrese and J. Cardy, J. Stat. Mech. (2007) P10004.

\bibitem{WorkSilva} A. Silva, Phys. Rev. Lett. {\bf 101}, 120603 (2008); A. Gambassi and A. Silva, Phys. Rev. Lett. {\bf 109} 250602 (2012).


\bibitem{Delft} W. M\"under, A. Weichselbaum, M. Goldstein, Y. Gefen and J. von Delft, Phys. Rev. B {\bf 85}, 235104 (2012)

\bibitem{Anderson} P.W. Anderson, Phys. Rev. Lett. {\bf 18}, 1049 (1967).

\bibitem{AffleckBarzykin} V. Barzykin and I. Affleck, Phys. Rev. B {\bf 61}, 6170 (2000).







\bibitem{BCCFF1} F. Lesage and H. Saleur, Phys. Rev. Lett. {\bf 80}, 4370--4373 (1998). 

\bibitem{BCCFF2} F. Lesage and H. Saleur, Nucl. Phys. B {\bf 520},  563--593(1998). 

\bibitem{Goold1} R. Dorner, J. Goold, C. Cormick, M. Paternostro, and V. Vedral, Phys. Rev. Lett. {\bf 109}, 160601 (2012). 

\bibitem{Gambassi} S. Sotiriadis, A. Gambassi and A. Silva, {\sl Statistics of the work done by splitting a one-dimensional condensate}, {\tt arXiv:1303.0782}.

\bibitem{Goold2} R. Dorner, S. R. Clark, L. Heaney, R. Fazio, J. Goold and V. Vedral, {\sl Extracting quantum work statistics and fluctuation theorems by single qubit interferometry}, {\tt arXiv:1301.7021}.

\bibitem{Campos} L. Campos Venuti and P. Zanardi, Phys. Rev. A {\bf 81}, 022113 (2010), L. Campos Venuti, N.T. Jacobson, S. Santra, and P. Zanardi, Phys. Rev. Lett. {\bf 107}, 010403 (2011).

\bibitem{EchoCFT1} J. Dubail and J.M. St\'ephan, J. Stat. Mech. (2011) L03002.

\bibitem{EchoCFT2} J.M. St\'ephan and J. Dubail, J. Stat. Mech. (2011) P08019.


\bibitem{GZ} S. Ghoshal and A.B. Zamolodchikov, Int. J. Mod. Phys. A {\bf 9}, 3841 (1994).


\bibitem{Smirnov} F. Smirnov, {\em Form-factors in completely integrable models of quantum field theory}, World Scientific (Singapore)

\bibitem{qu1} D. Iyer and N. Andrei, Phys. Rev. Lett. {\bf 109}, 115304 (2012). 

\bibitem{qu2} F.H.L. Essler, S. Evangelisti and M. Fagotti, Phys. Rev. Lett. {\bf 109}, 247206 (2012).

\bibitem{qu3}  J.-S. Caux and F.H.L. Essler, {\em Time evolution of local observables after quenching to an integrable model}, {\tt arXiv:1301.3806} (2013).

\bibitem{TbRef} A. Bransch\"adel, E. Boulat, H. Saleur and P. Schmitteckert, Phys. Rev. B {\bf 82}, 205414 (2010).



\bibitem{LesageSaleurFriedel} F. Lesage and H. Saleur, J. Phys. A: Math. Gen. {\bf 30} L457, (1997).


\end{thebibliography}
\end{document}